# Mapping Differential Protein-Protein Interaction Networks using Affinity Purification Mass Spectrometry

Prashant Kaushal,[1,2,3,8,*] Manisha R. Ummadi,[4,5,6,7] Gwendolyn M. Jang,[4,5,6,7] Yennifer Delgado,[1,2,3] Sara K. Makanani,[1,2,3] Sophie F. Blanc,[1,2,3] Decan M. Winters,[1,2,3] Jiewei Xu,[4,5,6,7] Benjamin Polacco,[4,5,6,7] Yuan Zhou,[4,5,6,7] Erica Stevenson,[4,5,6,7] Manon Eckhardt,[4,5,6,7] Lorena Zuliani-Alvarez,[4,5,6,7] Robyn Kaake,[4,5,6,7,*] Danielle L. Swaney,[4,5,6,7,*] Nevan Krogan,[4,5,6,7,*] and Mehdi Bouhaddou[1,2,3,8,9,*]

[1] Department of Microbiology, Immunology, and Molecular Genetics, University of California, Los Angeles, Los Angeles, CA, USA

[2] Institute for Quantitative and Computational Biosciences, University of California, Los Angeles, Los Angeles, CA, USA

[3] Molecular Biology Institute, University of California Los Angeles, Los Angeles, CA, USA

[4] Quantitative Biosciences Institute (QBI), University of California, San Francisco, San Francisco, CA, USA

[5] QBI Coronavirus Research Group (QCRG), University of California, San Francisco, San Francisco, CA, USA

[6] Department of Cellular and Molecular Pharmacology, University of California, San Francisco, San Francisco, CA, USA

[7] Gladstone Institute of Data Science and Biotechnology, J. David Gladstone Institutes, San Francisco, CA, USA

[8] Technical Contact

[9] Lead Contact

* Correspondence: bouhaddou@ucla.edu (M.B.), nevan.krogan@ucsf.edu (N.J.K.), danielle.swaney@gladstone.ucsf.edu (D.L.S.), robyn.kaake@gladstone.ucsf.edu (R.K.), prashantkaushal@ucla.edu (P.K.)

## SUMMARY

Proteins congregate into complexes to perform fundamental cellular functions. Phenotypic outcomes, in health and disease, are often mechanistically driven by the remodeling of protein complexes by protein-coding mutations or cellular signaling changes in response to molecular cues. Here, we present an affinity purification-mass spectrometry (APMS) proteomics protocol to quantify and visualize global changes in protein-protein interaction (PPI) networks between pairwise conditions. We describe steps for expressing affinity-tagged "bait" proteins in mammalian cells, identifying purified protein complexes, quantifying differential PPIs, and visualizing differential PPI networks. Specifically, this protocol details steps for designing affinity-tagged "bait" gene constructs, transfection, affinity purification, mass spectrometry sample preparation, data acquisition, database search, data quality control, PPI confidence scoring, cross-run normalization, statistical data analysis, and differential PPI visualization. Our protocol discusses caveats and limitations with applicability across cell types and biological areas.

For complete details on the use and execution of this protocol, please refer to Bouhaddou et al. 2023[1].

## HIGHLIGHTS

- Protocol for gene construct design, transfection, and affinity-purification mass spectrometry analysis.
- Mass spectrometry data quality control and protein-protein interactions scoring.
- Quantitative and statistical mass-spectrometry data normalization and analysis.
- Visualization of global changes in protein-protein interaction (PPI) networks.

## GRAPHICAL ABSTRACT

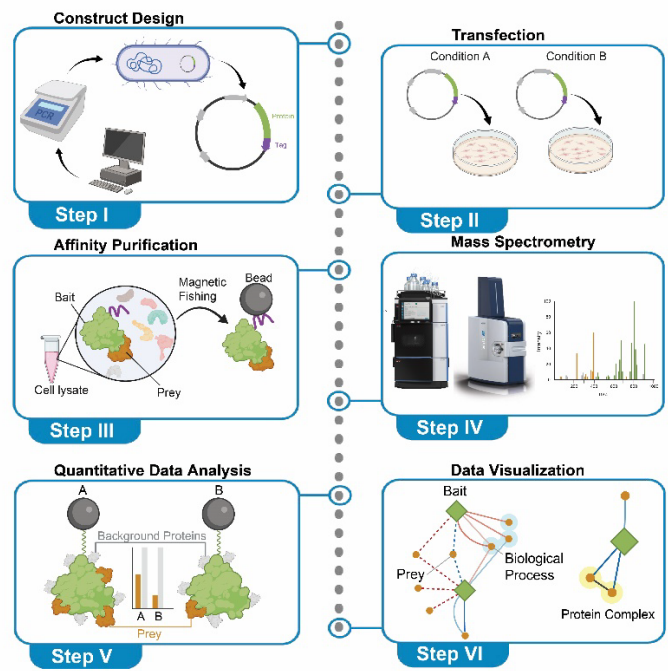

**BEFORE YOU BEGIN**

The protocol described in this paper has been previously employed to define changes in virus-host PPI networks between SARS-CoV-2 variants of concern (VOC) and their corresponding wave one (W1) viral protein forms using affinity purification-mass spectrometry (APMS) analysis in HEK293T cells. This approach can also be adapted to other cell types as long as target cells enable the introduction (i.e. transfection or transduction) and expression of a DNA construct.

a) **Institutional permissions**
   Timing: variable

   Researchers should obtain permission from the relevant institutions before conducting any BSL-2 level research with recombinant nucleic acid constructs and mammalian cell lines.

b) **Define groups for quantitative comparisons**
   Timing: variable

   Researchers should begin by defining two or more experimental conditions for comparison before defining changes in PPI networks. Our study compared SARS-CoV-2 W1 viral protein forms to their corresponding mutated variant of concern (VOC) forms. One may also choose to compare the same construct in the context of different cellular perturbations (e.g. drug administration), which may also regulate PPI changes. It is also important to define the negative controls that are relevant to your study; these typically include green fluorescence protein (GFP) and empty vector (EV) constructs.

**KEY RESOURCES TABLE**

| Reagent or Resource | Source | Identifier |
|---|---|---|
| | | |
| **Recombinant DNA** | | |
| SARS-CoV-2 VOC viral protein plasmids | Bouhaddou et al. 2023[1] | N/A |
| Stellar™ Competent Cells | Takara | Cat# 636766 |
| | | |
| **Antibodies** | | |
| Strep Tag Monoclonal Antibody | Thermo Scientific | Cat# MA5-17283 |
| | | |
| **Experimental models: Cell lines** | | |
| HEK293T cells | ATCC | Cat# CRL-3216 |
| | | |
| **Chemicals** | | |
| PolyJet™ DNA in vitro transfection reagent | SignaGen | Cat# SL100688 |
| S.O.C. Medium | Invitrogen | Cat# 15544034 |
| LB Broth (Powder) | Fisher | Cat# BP1427 |
| NucleoBond Xtra Midi kit | Macherey-Nagel | Cat# 740410.50 |
| Dulbecco's Modified Eagle Medium (DMEM) | Corning | Cat# 10-013-CV |
| Penicillin-Streptomycin (10,000 U/mL) | Gibco | Cat# 15140122 |
| Fetal bovine serum (FBS) | Gibco | Cat# 16140071 |
| Dulbecco's phosphate-buffered saline (DPBS), no calcium, no magnesium | Gibco | Cat# 14190144 |
| EDTA (0.5 M), pH 8.0, RNase-free | Invitrogen | Cat# AM9260G |

| NP-40 Surfact-Amps™ Detergent Solution | Thermo Scientific | Cat# 85124 |
|---|---|---|
| cOmplete™, Mini, EDTA-free Protease Inhibitor Cocktail | Roche | Cat# 11836170001 |
| PhosSTOP™ Phosphatase inhibitor tablets | Roche | Cat# 4906845001 |
| Bovine Serum Albumin Standard Pre-Diluted Set | Thermo Scientific | Cat# 23208 |
| Tris hydrochloride (Tris-HCl) | Invitrogen | Cat# 15506017 |
| NaCl (5 M), RNase-free | Invitrogen | Cat# AM9759 |
| Urea | Sigma Aldrich | Cat# U5128 |
| Dithiothreitol (DTT) | Sigma Aldrich | Cat# D0632 |
| Iodoacetamide (IAA) | Sigma Aldrich | Cat# I6125 |
| Sequencing-grade modified trypsin | Promega | Cat# V5111 |
| Trifluoroacetic acid (TFA), LC-MS grade | Thermo Scientific | Cat# 85183 |
| Formic acid (FA), LC-MS grade | Fisher Scientific | Cat# 85178 |
| Sep-Pak C18 1 cc vac cartridge (50 mg Sorbent per Cartridge) | Waters | Cat# WAT054955 |
| Water, Optima™ LC/MS Grade | Fisher Scientific | Cat# W64 |
| Acetonitrile, Optima™ LC/MS Grade | Fisher Scientific | Cat# A955 |
| Water with 0.1% Formic Acid (v/v) LC-MS grade | Thermo Scientific | Cat# LS118 |
| Acetonitrile with 0.1% Formic Acid (v/v) LC-MS grade | Thermo Scientific | Cat# LS120 |
| MagStrep "type3" XT beads | IBA lifesciences | Cat# 2-4090-010 |
| Buffer BXT (10x) | IBA lifesciences | Cat# 2-1042-025 |
| 6x Laemmli Sample Buffer | Thermo Scientific | Cat# AAJ61337AC |
| **Critical commercial assays** | | |
| Bradford protein assay kit | Thermo Scientific | Cat# 23200 |
| | | |
| **Software and algorithms** | | |
| MaxQuant | Cox and Mann, 2008[2] | https://www.maxquant.org/ |
| R statistical programming language | R CRAN | https://www.r-project.org/ |
| RStudio IDE | Posit | https://posit.co |
| MSstats | Choi et al. 2014[3] | https://bioconductor.org/packages/release/bioc/html/MSstats.html |
| artMS | Jimenez-Morales et al., 2023[4] | http://bioconductor.org/packages/artMS/ |
| Cytoscape | Shannon et al., 2003[5] | https://cytoscape.org/ |
| Spectronaut (Biognosys) | Bruderer et al., 2015[6] | https://biognosys.com/software/spectronaut/ |
| SAINTexpress | Teo et al., 2014[7] | https://saint-apms.sourceforge.net/Main.html |
| MiST | Jäger et al., 2011[8] | https://github.com/salilab/mist |
| CompPASS | Sowa et al., 2009[9] | http://pathology.hms.harvard.edu/labs/harper/Welcome.html |
| clusterProfiler | Wu et al., 2021[10] | https://bioconductor.org/packages/release/bioc/html/clusterProfiler.html |
| Adobe Illustrator | Adobe Inc. | https://adobe.com/products/illustrator |
| | | |
| **Deposited data** | | |
| APMS proteomics data | Bouhaddou et al. 2023[1] | PRIDE Project ID: PXD036798 |
| **Other** | | |
| **Resource** | **Source** | **Identifier** |
| Round bottom polystyrene test tube | Corning | Cat# 352001 |
| Tissue culture dishes (15-cm) | Fisher Scientific | Cat# FB012924 |
| Falcon™ 15 mL Conical Centrifuge Tubes | Fisher Scientific | Cat# 14-959-53A |
| Protein LoBind® Tubes (1.5 mL) | Eppendorf | Cat# 022431081 |
| Heating block | N/A | N/A |
| pH meter | N/A | N/A |
| pH strip | N/A | N/A |
| ThermoMixer C | Eppendorf | Cat# 2231001005 |
| ThermoTop | Eppendorf | Cat# 2231001048 |
| DynaMag-2 magnet | Invitrogen | Cat# 12321D |
| Nanodrop | N/A | N/A |
| Vacuum concentrator | N/A | N/A |
| Solid phase extraction vacuum manifold | N/A | N/A |
| Ultra high-pressure liquid chromatography (UHPLC) system | N/A | N/A |
| High-resolution mass spectrometer | N/A | N/A |

## MATERIALS AND EQUIPMENT

Note: Always use the highest available grade reagents for mass spectrometry analysis.

| **IP Buffer** | | |
|---|---|---|
| **Reagent** | **Final concentration** | **Volume** |
| Tris-HCl (pH 7.4) (1M) | 50 mM | 5 mL |
| NaCl (5 M) | 150 mM | 3 mL |
| EDTA (0.5 M) | 1 mM | 200 µL |
| Water | N/A | 91.8 mL |
| **Total** | **N/A** | **100 mL** |

Note: Store at 4°C for up to one month.

| **IP Lysis Buffer** | | |
|---|---|---|
| **Reagent** | **Final concentration** | **Volume** |
| Tris-HCl (pH 7.4) (1M) | 50 mM | 500 µL |
| NaCl (5 M) | 150 mM | 300 µL |
| EDTA (0.5 M) | 1 mM | 20 µL |
| NP-40 (10%) | 0.5% | 500 µL |
| Protease inhibitor | 1 tablet/10 mL | N/A |
| Phosphatase inhibitor | 1 tablet/10 mL | N/A |
| Water | N/A | ~8.5 mL |
| **Total** | **N/A** | **10 mL** |

Note: Always prepare fresh. Use immediately and discard the unused buffer.

| **IP Wash Buffer** | | |
|---|---|---|
| **Reagent** | **Final concentration** | **Volume** |
| Tris-HCl (pH 7.4) (1M) | 50 mM | 500 µL |
| NaCl (5 M) | 150 mM | 300 µL |
| EDTA (0.5 M) | 1 mM | 20 µL |
| NP-40 (10%) | 0.05% | 50 µL |
| Water | N/A | ~8.9 mL |
| **Total** | **N/A** | **10 mL** |

Note: Store at 4°C for up to one month.

| DMEM supplemented with FBS and Penicillin, Streptomycin mix | | |
|---|---|---|
| **Reagent** | **Final concentration** | **Volume** |
| DMEM | N/A | 445 mL |
| FBS | 10% | 50 mL |
| Penicillin-Streptomycin mix (10,000 U/mL) | 100 U/mL | 5 mL |
| **Total** | **N/A** | **500 mL** |

Note: Store at 4°C for one to two months. Check for mycoplasma contamination regularly.

| Denaturation-Reduction Buffer | | |
|---|---|---|
| **Reagent** | **Final concentration** | **Amount** |
| Urea | 2 M | 0.12 g |
| Dithiothreitol (1 M in water) | 1 mM | 1 µL |
| Tris-HCl (1 M, pH 8.0) | 50 mM | 50 µL |
| Water | N/A | ~945 µL |
| **Total** | **N/A** | **1 mL** |

Note: Always prepare fresh. Use immediately and discard the unused buffer.

| Alkylation solution | | |
|---|---|---|
| **Reagent** | **Final concentration** | **Amount** |
| Iodoacetamide | 0.1 M | 18.5 mg |
| Water | N/A | ~970 µL |
| **Total** | **N/A** | **1 mL** |

CRITICAL: Always prepare fresh and protect from light. Use immediately and discard the unused solution.

| **Trypsin solution** | | |
|---|---|---|

| Reagent | Final Concentration | Amount |
|---|---|---|
| Lyophilized Trypsin | 0.5 µg/µL | 20 µg |
| Acetic acid (50 mM) | 50 mM | 40 µL |
| **Total** | **N/A** | **40 µL** |

Note: Prepare aliquots for one-time use and store at -80°C for several months. Avoid multiple freeze-thaw cycles.

**STEP-BY-STEP METHOD DETAILS**

**Plasmid generation**
Timing: Between 1 week and 1 month

1. Obtain protein sequences of interest in a donor DNA plasmid. In our study[1], we performed site-directed mutagenesis on wave one (W1) SARS-CoV-2 gene plasmids to generate variant isoforms (Genscript Biotech).

Note: This approach is not limited to comparing changes in amino acid sequences but is amenable to the comparison of other conditions, such as cellular perturbations (e.g. drug treatment). In principle, this approach is used when minor changes in PPIs are expected, which is the case given small changes in protein sequence[11], such that one would expect quantitative changes in PPIs to occur (i.e. versus a binary response).

CRITICAL: Sequences were codon-optimized for expression in mammalian cells using either IDT codon-optimization (https://www.idtdna.com/codonopt) or Genscript Biotech tools.

2. Insert two copies of the Strep tag (2x-Strep) at either the N- or C-terminus of the protein (Table 1). Additionally, insert short linker sequences, containing glycines (G) and serines (S), between the 2x-Strep and the protein as well as between each Strep tag. We recommend starting by adding the tag on the C-terminal end to start; however, if the expression is not adequate as determined by Western blotting (evaluated below), we then attempt an N-terminal tag. See notes below for discussion on alternative affinity tags and tag location selection strategy.

**Table 1. Amino acid sequence of 2x-Strep tag for insertion at N- and C-terminus of protein.**

| 2x-Strep Tag Location | Amino Acid Sequence |
|---|---|
| N-term | MWSHPQFEKGGGSGGGSGGGSWSHPQFEKGGGGS |
| C-term | GGGGSWSHPQFEKGGGSGGGSGGGSWSHPQFEK* |

Note: Other affinity tags, including 3x-Flag[12] and 3xHA[13], may also be used. Tag selection should be optimized based on the protein of interest; some tags may work better than others for different proteins and cellular contexts. Another idea to consider is including an internal tag within the

protein sequence (as opposed to N- or C-term), which may have fewer adverse effects on protein function. However, the efficacy of these strategies must be evaluated on a case-by-case basis.

Note: To guide our tagging strategy, we used GPS-Lipid to predict protein lipid modification of the termini (http://lipid.biocuckoo.org/webserver.php), TMHMM Server v.2.0 to predict transmembrane/hydrophobic regions (http://www.cbs.dtu.dk/services/TMHMM/), and SignalP v.5.0 to predict signal peptides (http://www.cbs.dtu.dk/services/SignalP/). N-terminal signal peptides are often cleaved, precluding an N-terminal tag, which would be lost. The proximity of tags to modified residues and transmembrane domains may affect proper protein function or localization.

3. Clone DNA sequence containing protein of interest conjugated to an affinity tag into a plasmid with high expression in target cells. In our study, we cloned our DNA sequences into the lentiviral constitutive expression vector pLVX-EF1alpha-IRES-Puro (Takara Bio), driving constitutive protein expression under an eIF1α promoter in mammalian cells. Many alternative expression vectors exist and can be substituted here, even those with inducible expression (e.g. pLVX-TetOne-IRES-Puro).
   a. Perform transformation using Stellar competent *E. coli* cells per the manufacturer's instructions. Streak bacteria on agar plates. Grow at 37°C for 24-28 h.

Note: Different bacteria and/or constructs may require a lower temperature, such as 30°C. Consult the manufacturer's instructions.

   b. Pick individual colonies and confirm transformation efficacy via plasmid sequencing.
   c. Grow transformed bacterial cells in 10 mL sterile Lennox formulation (LB) broth containing 100 µg/mL ampicillin or other antibiotic corresponding to the encoded resistance gene overnight at 37°C with shaking. Scale up the culture volume according to the amount of plasmid required.

Note: Optimal shaking speed and temperature are construct-specific; thus, various conditions should be evaluated. Previously, we have shaken at speeds ranging from 140 rpm to 230 rpm, depending on the specific constructs, at 30-37°C.

   d. Perform plasmid purification per the manufacturer's instructions.

Note: We use the NucleoBond Xtra Midi kit (Macherey-Nagel) and recover ~1 mg of plasmid per construct from 200-250 mL of culture medium.

CRITICAL: When considering experimental design, it is important to include both a tagged GFP construct and an empty vector (no protein or tag) construct as negative controls. These controls are important when performing the SAINT PPI scoring following mass spectrometry analysis. Clone these into the same backbone expression plasmid as your proteins of interest.

**Transfection of affinity-tagged genes**
Timing: 1 week

4. Begin by determining the quantity of DNA needed for optimal expression of each protein. Each construct may require a different amount of DNA input in order to achieve adequate protein expression levels due to differences in translation efficiency and protein stability. To determine the optimal quantity of DNA for each construct,
   a. Transfect HEK293T cells at 80% confluency in a 24-well format with 0.1-0.4 µg DNA per construct, including GFP and empty vector (EV) control conditions.

Note: A starting range for the amount of DNA transfected in 24-well format should be determined by scaling down the DNA quantity based on the number of cells from a maximum of 15 ug total DNA transfected in a 15cm dish format.

   b. Determine the optimal quantity of GFP plasmid DNA needed such that cells appear green via fluorescence microscopy. In HEK293T cells, we aim to see greater than 80% of cells expressing GFP.
   c. Perform a Western blot with an anti-Strep-tag-II primary antibody. It is important to notice a clearly demarcated band at the appropriate molecular weight (see Figure 1 for an example of clearly demarcated bands) for each protein of interest. Use GFP as a positive expression control.

Note: Low protein expression, due to insufficient transfection efficiency, low construct expression, or high rates of protein degradation, will hinder the recovery of interacting proteins. Optimize DNA quantity to maximize protein expression in the absence of overt cytotoxicity.

   d. Adjust DNA input depending on Western blot results, ideally remaining between 0.1 and 0.4 µg DNA per well in a 24-well format. This will enable proper conversion of quantities to the 15-cm dish format.
   e. Following optimization, scale the amount of input DNA by the number of cells from a 24-well format to a 15-cm dish. For transfection in a 15-cm dish, keep the DNA amount less than 15 µg total (see Step 6).

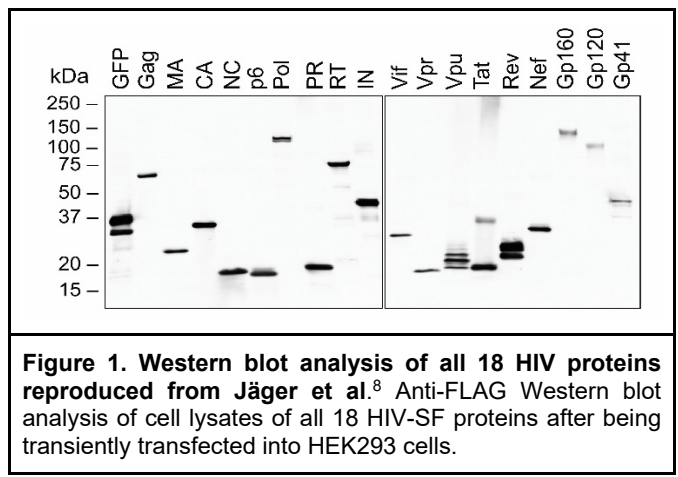

**Figure 1. Western blot analysis of all 18 HIV proteins reproduced from Jäger et al.**[8] Anti-FLAG Western blot analysis of cell lysates of all 18 HIV-SF proteins after being transiently transfected into HEK293 cells.

Note: Below, we describe how to perform cross-run normalization of prey intensities, which is critical for label-free APMS proteomics approaches. However, working to equalize bait expression between conditions is essential to achieve a reliable comparison between conditions. When comparing two baits, we recommend their expression falls within 2-fold; if outside of this range,

we recommend optimizing the transfection conditions (i.e. increasing/decreasing DNA concentration) to further equalize expression.

5. Once the optimal DNA quantity has been determined for each construct, seed 5-10 million HEK293T cells in 30 mL DMEM supplemented with 10% FBS and 1% penicillin-streptomycin in a 15-cm cell culture plate. Ensure the number of cells seeded results in 50-70% confluence at the time of transfection. In order to perform differential statistical analysis (below), it is important to include 3 separate replicates per experimental condition.

CRITICAL: Allow cells to reach approximately 50-70% confluence before proceeding with transfections, which typically occur in 16-24 hours. Transfecting with too few cells can lead to excessive cell death and/or insufficient protein recovery.

6. Sixteen to 24 hours after seeding, perform DNA transfection.
    a. Prepare Tube A. Combine plasmid DNA with serum-free DMEM.
        i. Aliquot each DNA plasmid in separate 2 mL tubes. Use a separate tube for each replicate. Use 15 µg total plasmid per 15-cm dish. Determine the mass of plasmid DNA required for optimal expression (above), then add empty vector DNA (i.e. backbone construct) to a total of 15 µg DNA (gene + empty vector = 15 µg total).
        ii. Add DMEM media to each tube with aliquoted DNA to a total of 500 µL.
        iii. Vortex each tube briefly to mix.

    b. Prepare Tube B. Prepare mastermix of PolyJet transfection reagent with serum-free DMEM.
        i. Vortex PolyJet transfection reagent well before use. Aliquot PolyJet into a tube large enough to fit 500uL multiplied by the number of total samples (i.e. replicates) as a single mastermix will be used for all samples. Use 3 µL of PolyJet transfection reagent per 1 µg of plasmid (e.g., for 15 µg of plasmid, use 45 µL of PolyJet).
        ii. Add serum-free DMEM to PolyJet to a total of 500 µL per sample. Scale PolyJet and serum-free DMEM mix accordingly (Table 2).
        iii. Vortex the tube briefly to mix.

**Table 2. PolyJet / DMEM solution volume to create a master mix (Tube B) for 15 µg DNA.**

|  | Single plate | For three plates | Total + ~10% error |
|---|---|---|---|
| PolyJet (µL) | 45 | 135 | 148 |
| DMEM (µL) | 455 | 1365 | 1502 |
| Total (µL) | 500 | 1500 | 1650 |

Note: When performing multiple transfections, create a master mix of Tube B in a larger tube (e.g. 15 mL Falcon tube).

   c. Add 500 µL of Tube B (PolyJet/DMEM) to each Tube A (DNA/DMEM), resulting in a total volume of 1 mL per tube. Invert each tube 3-4 times to mix gently. Incubate for 15 min at room temperature (RT) to allow PolyJet-DNA complexes to form.

d.  Following incubation, carefully add transfection complexes dropwise to cells. Mix well by carefully tilting dishes back-and-forth and side-to-side several times.
   e.  Incubate cells for 48 hours post transfection at 37ºC/5% $CO_2$ prior to harvesting for affinity purification.

Note: Alternative transfection reagents, such as Lipofectamine 3000 (Thermo Scientific), can also be used. We have found that PolyJet provides efficient transfection and adequate protein expression with our SARS-CoV-2 constructs in HEK293T cells and is a cost-effective solution for large-scale experiments. Alternative transfection reagents may be more effective for other cell types and plasmids. Transfection conditions should be optimized on a case-by-case basis.

**Cell harvest**
Timing: 30 min - 4 hours

Note: The cell harvest procedure described was developed for HEK293T cells, which detach from the cell culture dish in 10 mM EDTA. Cell harvest may need to be performed differently for different cell types. We recommend avoiding the use of trypsin to detach cells from the culture plate since this will result in the loss of the peptides from proteins on the cell surface. If using adherent cells that do not detach under 10mM EDTA conditions, consider adding the lysis buffer directly to the dish and scraping the cells before transferring them to a cold 1.5 mL tube.

7. Carefully aspirate the supernatant from 15-cm dishes using a vacuum line fitted with a P200 pipette tip. Add 10 mL Dulbecco's phosphate-buffered saline (DPBS), without calcium and magnesium, to each plate, supplemented with 10 mM EDTA.

8. Incubate dishes for 5-10 min at RT until cells mostly detach from the monolayer. Gently shake and tap the side of the dish to detach any remaining cells.

9. Transfer each plate of detached cells to individual cold 15 mL Falcon tubes on ice. Store the cell suspension on ice until the final step of the harvest. Centrifuge the cell suspension at 400xg for 5 min at 4°C. Carefully aspirate the supernatant using a vacuum line fitted with a P200 pipette tip.

10. Wash the cells by gently resuspending the cell pellet in 10 mL DPBS. Centrifuge the cell suspension at 400xg for 5 min at 4°C. Carefully aspirate the supernatant using a vacuum line fitted with a P200 pipette tip. Repeat the wash step once more.

11. Using a P1000 pipet tip, resuspend the cell pellet in 1 mL ice-cold DPBS and transfer cell suspension to a cold 1.5 mL protein lo-bind tube on ice. Centrifuge at 400xg for 5 min at 4°C in a microcentrifuge. Carefully aspirate the supernatant using a vacuum fitted with a P200 pipette tip. Ensure that all DPBS is removed from the surface of the cell pellet.

CRITICAL: Be careful to not accidentally aspirate the cell pellet. Fitting a P200 pipette at the end of the vacuum line should enhance precision of DPBS aspiration.

12. Snap freeze cell pellets immediately on dry ice or liquid nitrogen and store at -80°C until ready to proceed to subsequent steps.

Note: Once snap frozen on dry ice and placed at -80ºC, cell pellets can be stored for several months before subsequent steps without substantial loss in protein integrity.

**Strep-tag affinity purification**
Timing: 2 days

13. Thaw frozen cell pellets on ice for 15-20 min and resuspend in 1 mL IP Lysis Buffer (IP Buffer supplemented with 0.5% NP-40 and protease/phosphatase inhibitors).

Note: A freeze-thaw cycle can be added here in order to improve lysis efficiency. Lysates should be frozen on dry ice for at least 10 min (up to overnight) and then thawed on ice for 15-20 min. Multiple freeze-thaw cycles can be implemented to improve lysis efficiency further, up to a maximum of three cycles. However, performing too many additional freeze-thaw cycles may also increase protein degradation and reduce protein recovery.

14. Centrifuge at 13,000xg for 15 min at 4°C to clarify the lysate and pellet debris. Protein will remain in the supernatant.

Optional: To save samples for Western blot analysis, aliquot 50 μL of clarified lysate and dilute with 4-6x Laemmli Sample Buffer (SB). Store in PCR strip tubes or 1.5 mL protein LoBind tubes and proceed to Western blot analysis. Optionally, freeze lysates in the Laemmli SB for several weeks at -80°C until ready to run the Western blot.

15. Prepare MagStrep "type3" XT beads.
    a. Pipet beads up and down using a wide-bore pipet tip to resuspend stock bead slurry.
    b. Prepare one 1.5 mL protein LoBind tube per sample.
    c. Aliquot 30 μL slurry into each tube.

Note: Beads settle down quickly, resulting in unwanted sample-to-sample variability. Pipetting up and down to mix the slurry after every few samples ensures consistent aliquoting of the beads across the samples.

    d. Wash MagStrep "type3" XT beads 2x with 1 mL IP Wash Buffer (IP Buffer with 0.05% NP-40) using a magnetic rack. Specifically, place beads on a magnetic rack and remove the solution, leaving the beads behind. Add 1 mL IP Wash Buffer, remove tubes from the magnetic rack, and vortex briefly to mix. Place tubes back on the magnetic rack, remove the solution, leaving beads behind, and continue to the next wash.
    e. After the second wash, resuspend beads in 0.3 mL IP Buffer.

16. Add the remaining protein lysate (1 mL minus any put aside for Western blot) to the beads and incubate for 2 h at 4°C on an end-over-end tube rotator.

17. Centrifuge at 600xg for 30 sec to pellet the beads. Place tubes on the magnet. Discard the supernatant and resuspend the beads in 1 mL of IP Wash Buffer. Rotate the tubes on an end-over-end tube rotator for 5 min at 4°C. Repeat this step two additional times. After the wash, resuspend beads in 1 mL of IP Buffer.

Optional: Protein can also be eluted from the beads to check the expression of the target step-tagged protein and any interactors by performing a silver stain following SDS-PAGE. To do this,
   a. Move 200 µL (20%) of mixed bead slurry to a new 1.5 mL protein LoBind tube.
   b. Collect beads on the magnetic rack and discard the solution, leaving beads behind.
   c. Add 30 µL of 1X Buffer BXT (dilute 10X Buffer BXT 1:10 with water) and gently agitate on an electronic shaker (e.g. Eppendorf ThermoMixer C) for 30 min at RT.
   d. Collect beads on the magnetic rack and transfer eluates to a fresh 0.5-1.5 mL protein LoBind tube.
   e. Add Laemmli SB and proceed with SDS-PAGE and silver stain. Optionally, freeze eluates in Laemmli SB for several weeks at -80°C.

18. To perform an "on-bead digest," digesting proteins bound to the beads,
   a. Briefly collect beads on the magnetic rack and discard the solution, leaving beads behind.
   b. Add 50 µL Denaturation-Reduction Buffer and incubate for 30 min at 37°C, 1100 rpm, on an electronic mixer (e.g., Eppendorf ThermoMixer C with a heated lid like ThermoTop).
   c. Add 1.5 µL of alkylation solution (final concentration of IAA is 3 mM) and incubate for 45 min at RT, 1100 rpm on the electronic mixer.

CRITICAL: Protect samples from light during this step by covering with aluminum foil.

   d. Add 1.6 µL of 0.1 M DTT (final concentration is 3 mM) and incubate for 10 min at RT, 1100 rpm on the electronic mixer.
   e. Add 15 µL of 50 mM Tris-HCl (pH 8.0) to offset evaporation to each sample. Skip this step if using a heated lid, which should prevent condensation.
   f. Add 1.5 µL of stock trypsin solution (0.5 µg/µL) and incubate for 4-6 h at 37°C, 1100 rpm on the electronic mixer.
   g. Add an additional 0.5 µL of stock trypsin solution and incubate at 37°C for 1-2 h at 1100 rpm on the electronic mixer.
   h. Briefly collect beads on the magnetic rack and transfer the digest in the supernatant to a new protein LoBind tube.
   i. Resuspend the beads in 50 µL of 50mM Tris-HCl (pH 8.0) and pool with the supernatant from the previous step.

Note: A KingFisher Flex (KFF) purification system can automate the protocol steps. For affinity purification, place KFF in a cold room and allow it to equilibrate to 4°C overnight before use. Use a slow mix speed and the following mix times: 30 s for equilibration and wash steps, 2 h for binding, and 1 min for final bead release. Use three 10-second bead collection times between all steps. A KFF protocol file in both pdf and bdz formats is provided in Supplementary Information (SI).

19. To purify samples prior to mass spectrometry analysis, use one C18 Sep-Pak cartridge containing 50 mg sorbent (suitable for up to 500 µg of peptides),

a. Add 10% TFA in water to the digested peptides to a final concentration of 0.5% and check pH with a pH strip. It should be a pH less than three. Continue to add 10% TFA as needed to reach the desired pH.
   b. Place C18 Sep-Pak columns on a solid phase extraction vacuum manifold.
   c. Activate the C18 Sep-Pak column with 1 mL of 80% ACN (v/v), 0.1% TFA in HPLC water. Discard the flowthrough.
   d. Equilibrate the column with 1 mL of 0.1% TFA in HPLC water. Repeat two additional times. Discard the flowthrough.
   e. Add the sample to the column. Discard the flowthrough.

Optional: Flowthrough can be collected and stored to assess sample loss during desalting.

   f. Wash peptides with 1 mL of 0.1% TFA in HPLC water. Repeat two additional times. Discard the flowthrough.
   g. Elute bound peptides with 400 µL of 50% ACN (v/v), 0.25% formic acid in HPLC water. Repeat this step an additional time.
   h. Dry peptides in a vacuum concentrator.

Note: Do not allow the Sep-Pak column sorbent to dry at any step. It may significantly decrease the peptide recovery.

Pause point: Dried peptides can be stored at -80°C for several months.

**Mass Spectrometry data acquisition and database search**
Timing: 1-2 days

20. Resuspend dried peptides in 50 µL of 0.1% formic acid in water and estimate the peptide concentration using a Nanodrop.

21. Use an ultra-high-pressure liquid chromatography (LC) system paired with a mass spectrometer (MS) to separate peptides on a reverse-phase C18 column on a gradient of mobile phase A (100% water, 0.1% formic acid) and mobile phase B (80% ACN, 0.1% formic acid). In our study, we used an Easy-nLC 1200 ultra-high-pressure liquid chromatography system (Thermo Fisher Scientific).
   a. Inject approximately 500 ng sample on a reverse phase column (25 cm length x 75 µm i.d.) packed with ReprosilPur 1.9 µm C18 particles.

Note: The amount of peptides to inject is MS instrument-dependent, as some instruments may require more or less peptides to achieve the same sensitivity.

   b. Equilibrate the column with mobile phase A and separate peptides using a gradient of mobile phase B from 2% to 7% over 1 min, followed by an increase to 36% B over 53 min, then hold at 95% B for 13 min, then reduce back down to 2% B for 11 min at a flow rate of 300 nL/min.

Note: The LC gradient may require slight adjustments in an LC and MS instrument-specific manner.

Note: A trap-and-elute setting can be used in place of direct injection to perform an additional online sample clean-up.

22. For the mass spectrometry analysis of peptides, use a high-resolution mass spectrometer in either data-dependent acquisition (DDA) or data-independent acquisition (DIA) mode. In our study[1], we used an Orbitrap Exploris 480 mass spectrometer (Thermo Fisher Scientific) coupled to an Easy-nLC 1200 ultra-high-pressure liquid chromatography system (Thermo Fisher Scientific) with a Nanospray Flex nanoelectrospray source (Thermo Fisher Scientific).
    a. For data-dependent acquisition (DDA) mode, perform a full scan over an m/z range of 300–1500 in the Orbitrap at >50,000 resolving power with an AGC target of 1e6 with an RF lens setting of 40%. Set dynamic exclusion to 45 s and exclusion width to 10 ppm. Fragment top 20 peptides, within charge state 2-6, with high-energy collision dissociation (HCD) or collision-induced dissociation (CID) at 20 MS/MS scans per cycle and a resolving power of 17,500.
    b. For acquiring data in data-independent acquisition mode (DIA), perform MS scan at 60,000 resolving power over a scan range of 350–1100 m/z at a normalized AGC target of 300% and an RF lens setting of 40%. Perform DIA scans at 15000 resolving power, using 20 m/z isolation windows over 350–1100 m/z at a normalized HCD collision energy of 30%.

Note**:** Aliquots from each set of three biological replicates can be pooled and acquired in DDA mode to build a spectral library. Alternatively, DIA data can be searched using an in silico-generated spectral library.

Note: Other high-resolution mass spectrometers designed for proteomics can be substituted here for DDA and DIA analyses.

23. To search the raw MS data,
    a. For DDA data analysis, analyze raw files using MaxQuant[2] with default settings (or another search engine of your choice) to search the data against relevant proteomes. In our study, we used *Homo sapiens* and SARS-CoV-2 proteomes.
    b. For DIA data analysis,
        i. For library-based search, first, build an experiment-specific spectral library from the DDA data using Spectronaut[6] or DIA-NN[14] (or another search engine of your choice). Then, search the DIA data using the spectral library generated in the previous step. In Spectronaut, use the BGS Factory setting (default) workflow.
        ii. For library-free search, build an in-silico spectral library. In Spectronaut, use the directDIA workflow with the default settings.
        iii. For either analysis, remove cross-run normalization (which will be performed later in MSstats) and imputation of missing values.

Note: For all searches, set methionine oxidation as a variable modification and carbamidomethyl cysteine as a static modification. Filter results to a final 1% false discovery rate (FDR) at the peptide spectrum match (PSM), peptide, and protein levels.

**Data quality control**

Timing: 1 day

24. Evaluate peptide intensity correlations between replicates of the same condition. We recommend using a square matrix heatmap-based visualization approach, where each replicate populates the rows and columns in the same order. Each cell will be colored and labeled according to the Pearson's r correlation coefficient. Correlation coefficients between biological replicates should be greater than 0.8 to preserve them; if they fall below this number, discard the problematic replicate prior to subsequent analyses.

25. Evaluate peptide intensity pattern consistency between replicates using principal components analysis (PCA). Visualize the first and second principal components on the x and y-axes, respectively. If a replicate does not appear to cluster with the others, discard this replicate.

26. Evaluate differences in bait expression by comparing the abundance of bait peptides across runs. To do this, first, for each pair of conditions, identify a set of bait peptides that are detected across all biological replicates and sum their intensities. Ensure that the resulting summarized bait protein intensities are within 2-fold between each pair of conditions being compared. If bait expression levels are greater than 2-fold between conditions, we recommend optimizing the transfection conditions (i.e. increasing/decreasing DNA concentration) to further equalize expression and redoing the experiment.

27. Evaluate the sum of all peptide intensities for each sample. Some samples may result in higher overall peptide intensities. Here, consistency should be evaluated between the biological replicates. If there is greater than a 2-fold difference between biological replicates of the same condition, discard the outlying replicate prior to subsequent analysis.

28. Evaluate peptide and protein counts per sample. If a sample possesses a greater than 2-fold difference in peptide or protein counts relative to other replicates of the same condition, discard that replicate prior to subsequent analyses.

Note: We used the artMS R package[4] to generate figures that were interpreted to perform quality control analyses.

**Scoring protein-protein interactions**
Timing: 1 day

29. Perform PPI scoring using SAINTexpress[7]. This scoring algorithm assesses the abundance of prey relative to the negative controls (i.e., empty vector and GFP). We consider a prey significant if its false discovery rate (i.e., Benjamini & Hochberg false discovery rate "BFDR") is less than 0.05.

30. Additionally perform PPI scoring using MiST[8] or compPASS[9]. This scoring algorithm assesses prey specificity across the baits in your experiment. For example, if a prey binds to many baits in your sample, it is likely part of the background. This pattern of non-specific binding results in a score of lower confidence. For MiST, we consider a prey to be significant if the MiST score is greater than 0.7. For compPASS, we convert the WD score into a percentile and consider a prey to be significant if the WD percentile is greater than 98%.

Note: The MiST score threshold should be evaluated separately for each dataset as it is affected by the dataset size. In the updated MiST algorithm (version 1.5), the MiST score now automatically scales to dataset size.

Note: Traditionally, we have used MiST scoring for virus-host interaction datasets and compPASS for host-host interaction datasets.

Note: Specificity scoring algorithms like MiST and compPASS work best when many different conditions are included, since they work by comparing across different conditions. If only two conditions are included (the minimum), consider using SAINTexpress alone. Furthermore, MiST and compPASS work best when a dataset includes different bait proteins; in our study, this equated to including all SARS-CoV-2 proteins. Include exclusionary criteria for similar baits (including for pairs of conditions being compared if the baits contain high sequence similarity). This ensures specificity algorithms will not adversely penalize preys that are discovered across similar baits, which is biologically expected.

31. To create our final set of high confidence preys, we use a combination of abundance (e.g. SAINTexpress) and specificity (e.g. MiST or compPASS) cutoffs. In the past, we have required MiST>0.7 (or compPASS WD percentile > 98%) & SAINTexpress BFDR<0.05. We additionally require that each prey possesses an average spectral count (among biological replicates) of at least 2. The value of specific thresholds used should be tailored to each dataset, especially for their ability to identify any known interactors balanced against their inclusion of common contaminants or non-specific binders[8].

**Quantitative data analysis**
Timing: 1 day

32. Perform quantitative statistical analysis of protein abundance changes between conditions using peptide ion fragment data from the full datasets (not yet filtered by high-confidence preys).
    a. Export peptide ion fragment data from your search algorithm of choice and analyze using MSstats workflows. MSstats has several built-in pipelines to convert evidence files from MaxQuant and others to MSstats format.
        i. Define contrasts, which denote comparisons between conditions of interest.
        ii. Convert MaxQuant evidence files to MSstats format using MaxQtoMSstatsFormat with settings:
            1. D = ''Leading.razor.protein'', useUniquePeptides = FALSE, summaryforMultipleRows = sum, removeFewMeasurements = FALSE, removeOxidationMpeptides = FALSE, and removeProtein_with1Peptide = FALSE.
        iii. Run the dataProcess function with featureSubset = ''all'', normalization = "equalizeMedians'', MBimpute = FALSE, and summaryMethod = ''TMP''. In essence, this performs a cross-run normalization by median equalization, does not impute missing values, and summarizes multiple peptide ion or fragment intensities into a single intensity for their protein group.
        iv. Perform statistical tests of differences in intensity between conditions of interest (defined by contrasts, above) using defaults for MSstats for adjusted P values, even in cases of n = 2. By default, MSstats uses the Student's t-test for P value

calculation and the Benjamini–Hochberg method of FDR estimation to adjust P values. This analysis results in log2 fold changes (log2FC) and p-values per interaction between corresponding mutant and wave 1 baits.

Note: As described above, we suggest starting by performing MSstats normalization using global median equalization, no imputation of missing values, and median smoothing to summarize multiple peptide ion or fragment intensities into a single intensity for their protein group. However, an alternative approach is to normalize by bait expression. Global median equalization essentially normalizes to the background, correcting for differences in overall peptide intensities injected into the mass spectrometer for each run, important to correct for variability in input quantity and sample handling. Bait-based normalization assumes prey intensities scale linearly with bait abundance, an assumption originating from the law of mass action. However, bait abundance could be at saturating intensities, such that further increases in bait abundances do not correspond to increases in prey abundances. Thus, the choice of normalization should be evaluated and applied in a dataset-specific fashion. Typically, we do not expect large differences in protein-protein interactions between conditions[15]; for example, we often compare proteins with a single amino acid change. In such cases, the prey distribution of $log_2$ fold changes should be centered around zero, without a major skew in either direction (i.e., all or most preys increase or decrease). If this is observed, we recommend altering the normalization approach. There are exceptions to cases like these, if, for instance, all increasing preys are known to be part of a complex.

Note: If normalizing by bait abundance, we recommend using the overlapping set of bait peptides identified in all replicates of the two conditions being compared so as not to artificially skew bait, and the resulting prey, abundances. Additionally, we recommend performing normalization using a custom-build pipeline prior to running MSstats (with normalization turned off).

Note: Although cross-run normalization is a critical aspect of label-free APMS-based proteomics, working to equalize bait expression between conditions is essential to achieve a reliable comparison between conditions. When comparing two baits, we recommend their expression falls within 2-fold; if outside of this range, we recommend optimizing the transfection conditions (i.e. increasing/decreasing DNA concentration) to further equalize expression.

33. To define significantly different protein-protein interactions, or "differential protein-protein interactions",
    a. Define differential interactions based on two criteria:
        i. The prey must be a high-confidence interaction in either condition being compared (see scoring thresholds above) AND
        ii. The prey must be changing in abundance between the conditions being compared with an absolute value log2FC>1 and p<0.05.

Note: Thresholds for differential interactions should be tailored for each dataset.

**Data visualization**
Timing: 1 day - 1 week

34. Visualize differentially interacting proteins as a heatmap,

a. Create a heatmap with distinct bait comparisons (i.e. mutant versus wild-type) along the rows and preys along the columns (see Figure 2A). The heatmap can be made to include all preys that are high confidence in either condition, many of which will not significantly change between conditions; however, we only include preys that are significantly differentially interacting (based on criteria outlined in Step 37) for at least one of the comparisons tested. We recommend coloring each cell with the log2 fold change between conditions for each prey or gray if not detected in either condition. Visually indicate if a prey is only detected in one of the conditions (i.e. mutant only or wildtype only), thus possessing an infinite log2 fold change; we annotate these using white hatches. Lastly, annotate significantly different interactions (i.e. $p<0.05$); we have used a black bounding box to indicate $p<0.05$ (see Figure 2A).

35. Visualize differentially interacting proteins as a network,
    a. Create two text files, one called "edges" and one called "nodes". The edges file contains the pairwise interactions between bait and prey proteins, as well as any other edge annotations. The node file contains a list of proteins and annotation as to whether they are a bait or a prey.
    b. To make the edges file,
        i. Extract a network of differentially interacting proteins and their baits. Typically this takes the form of a table with four columns: (1) bait, (2) prey, (3) log2 fold change, (4) p-value, and (5) data source. The log2 fold change refers to the magnitude of change of prey abundance in the affinity purification between conditions. Data source refers to either "APMS" or "CORUM", where APMS refers to edges derived from the experimental mass spectrometry measurement and CORUM refers to edges added from the CORUM database[16] (see below).
        ii. Add edges from the CORUM database. First, download a table of interactions from the online resource[16]. Next, merge protein complex interactions from CORUM that exist between any two preys bound to the same bait.

Note: To simplify network visualization, we typically draw CORUM edges between preys that are bound to the same bait and not between preys bound to different baits. However, it is also possible to include edges between preys bound to different baits. Furthermore, additional protein-protein interaction databases can be integrated; we recommend CORUM because it contains high-confidence protein complexes with well-studied functions.

    c. To make the nodes file,
        i. Collapse the unique proteins in the bait and prey columns in the edges table into a column called "nodes". Add an additional column called "is_bait", which is given a TRUE if the protein is a bait and FALSE if the protein is a prey.

    d. Visualize differential interaction network using Cytoscape[5].
        i. Import edges file using the "Import Network from File System" button. Select source and target nodes as bait and prey columns, respectively. All other columns will default to "edge attributes".
        ii. Import nodes table using the "Import Table from File" button. Make sure to import into the loaded network.

  iii. Set edge thickness or color to the data source or log2 fold change columns. Optionally you could set the thickness of the line proportional to the -log10(p).
  iv. Change the node shape according to the "is_bait" column in the nodes file. We set baits to be diamonds and preys to be circles. We additionally make bait nodes bigger than prey nodes.
  v. Arrange nodes manually in a visually aesthetic manner. One strategy is to start with a Circular Layout and arrange bait-prey interactions in a circular format (see Figure 2).

Note: Change prey protein labels to be centered to the outside of the node. These can be adjusted in graphics software (e.g. Adobe Illustrator, see below) to not overlap any other nodes or edges.

 e. Once nodes are arranged in Cytoscape, export as PDF and import into Adobe Illustrator for final aesthetic adjustments.
  i. Add colorful circles for protein complexes and biological processes. Protein complexes are evident from CORUM edges incorporated into the network. Biological process terms must be manually refined from a GO Biological Process gene overrepresentation enrichment analysis using the clusterProfiler[10] package in R, for which there are several online tutorials available. Specifically, perform the enrichment analysis on the group of preys from each bait separately. Next, manually group preys into shared biological processes. We recommend only annotating preys with a biological process term if their primary function is associated with said term. To accomplish this, it is important to read about the known function(s) of all genes prior to finalizing an annotation. If a gene has multiple known functions or its connection to a certain biological process is unclear, avoid annotating this gene. We recommend annotating biological processes within the preys for each given bait, and not between preys of different baits, which we feel simplifies the interpretation of the results. However, this decision is dataset dependent and should be decided on a case-by-case basis.

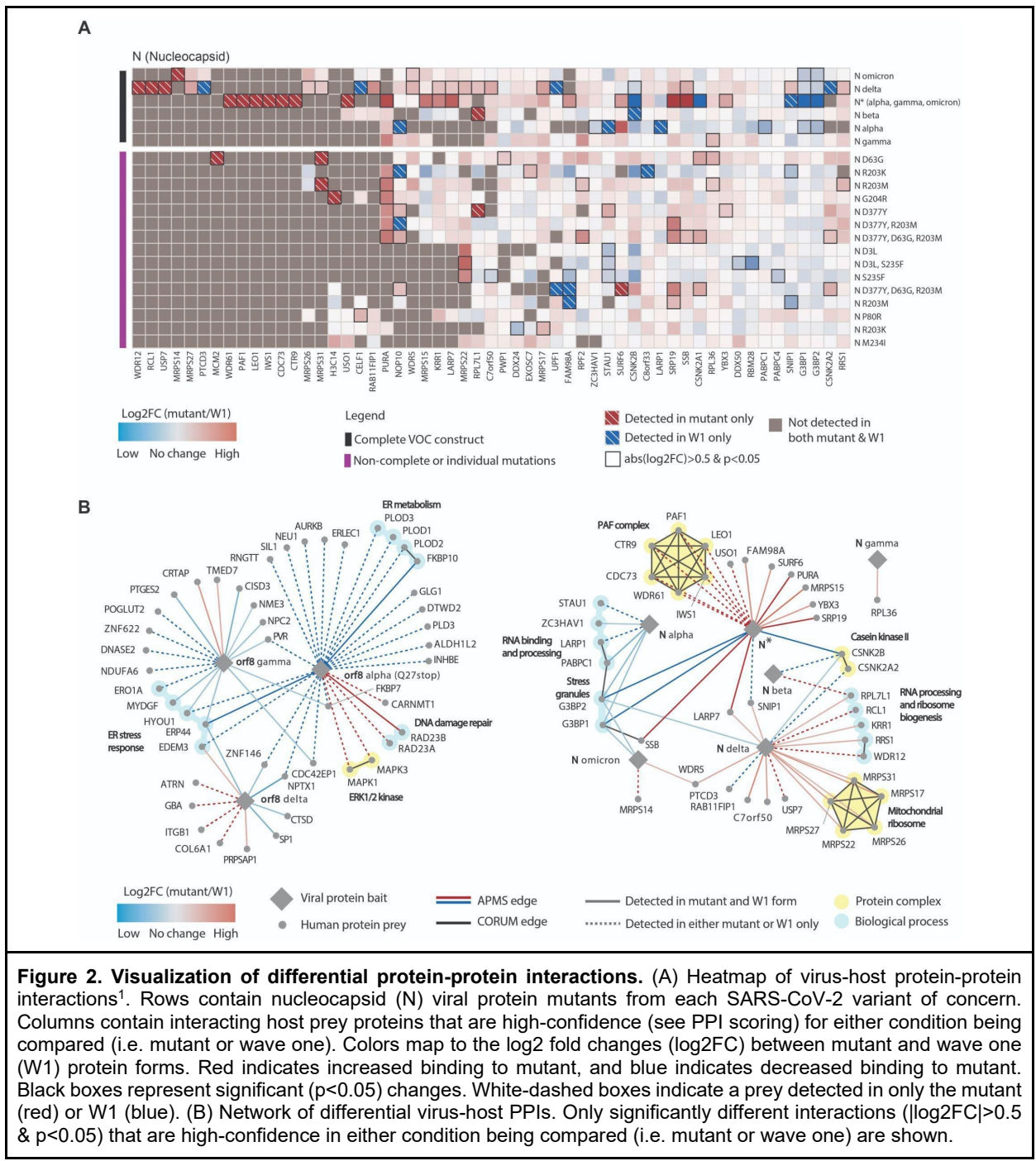

**Figure 2. Visualization of differential protein-protein interactions.** (A) Heatmap of virus-host protein-protein interactions[1]. Rows contain nucleocapsid (N) viral protein mutants from each SARS-CoV-2 variant of concern. Columns contain interacting host prey proteins that are high-confidence (see PPI scoring) for either condition being compared (i.e. mutant or wave one). Colors map to the log2 fold changes (log2FC) between mutant and wave one (W1) protein forms. Red indicates increased binding to mutant, and blue indicates decreased binding to mutant. Black boxes represent significant ($p<0.05$) changes. White-dashed boxes indicate a prey detected in only the mutant (red) or W1 (blue). (B) Network of differential virus-host PPIs. Only significantly different interactions (|log2FC|>0.5 & $p<0.05$) that are high-confidence in either condition being compared (i.e. mutant or wave one) are shown.

## EXPECTED OUTCOMES

This protocol provides researchers with an APMS proteomics pipeline integrated with a quantitative and statistical computational analysis to quantify and visualize differential PPIs. The expected outcome is a heatmap and network visualization of significantly changing protein complexes. In our original study, we used this protocol to generate PPI networks for SARS-CoV-2 VOCs and showed that compared to other VOCs, Omicron BA.1 possessed altered regulation of interferon stimulated genes (ISGs) relative to other VOCs, which correlated with altered SARS-CoV-2 Orf6-nuclear pore interaction affinities. However, this pipeline possesses general applicability to the comparison of pairs of proteins from any organism or between specific conditions/treatments of interest.

**LIMITATIONS**

A limitation of this protocol is that it captures high-affinity interactions representing highly stable protein complexes rather than transient interactions, such as kinase-substrate interactions. Proximity labeling approaches, like TurboID,[17] are better suited to capture transient interactions. Additionally, this approach does not identify whether a PPI is direct or indirect. For example, interactions may occur through a protein or nucleic acid intermediate, such as RNA. Methods to remove RNA-dependent interactions can be attempted in such cases (e.g. benzonase). Moreover, cell line/type variability in transfection/construct expression efficiency may render this approach difficult to implement in specific cell types/cell lines. In place, a viral transduction or electroporation approach to deliver the construct(s) may be utilized to enhance cell distribution.

**TROUBLESHOOTING**

Problem 1

Insufficient target protein expression.

Potential solution

- Ensure the plasmid is codon optimized.
- Confirm the bait identity by plasmid sequencing.
- Always use DNase-free tubes or reagents and maintain sterile conditions while purifying plasmid from bacterial culture.
- While preparing the plasmid/transfection reagent mixture (other than PolyJet), the tubes should be gently mixed by inverting them. Vigorous shaking may lead to plasmid disintegration and low transfection efficiency.
- Optimize the total plasmid amount and plasmid-to-transfection reagent ratio. Excess plasmid and/or transfection reagent may have cytotoxic effects, reducing transfection efficiency.

Problem 2

Insufficient enrichment of Strep-tagged target proteins.

Potential solution

- Before binding the sample to the MagStep 'type-3' XT beads, clear the cell lysate by centrifugation to remove any cell debris. Avoid re-use the MagStep 'type-3' XT beads.
- Increase the starting cell lysate amount and optimize the protein-to-bead ratio.

Problem 3

High number of non-specific interactors in the background.

Potential solution

To minimize the carryover of background proteins between washes, increase the beads' washing steps with the IP Wash Buffer and change the tubes in between.

Problem 4

No/low peptide hits for target protein in the mass spectrometry data.

Potential solution

First, check the purity of the sample via SDS-PAGE followed by coomassie or silver stain. The enriched sample should have a relatively larger target protein band than the empty vector sample.

- Use LC-MS-grade chemicals for the proteomics sample preparation and run a quality check (like HeLa protein digest) to ensure good LC-MS performance before analyzing the sample.
- Depending on the LC and MS instrument configuration, increase the total injected peptide amount for adequate sequence coverage.
- Use DIA-based data acquisition in place of DDA to minimize missing data values.

Problem 5

Network visualization appears chaotic.

Potential solution

A few simple strategies can improve the aesthetic quality and interpretability of a network diagram are detailed below.

- Move bait and prey names from the center of the node to open black space adjacent to the node (see Figure 2B). This can be finalized within a software such as Adobe Illustrator involving manually moving each label such that it does not directly touch any edge, node, or other label.
- Incorporate additional prey-prey protein-protein interaction data from high confidence databases with manually curated complexes, such as CORUM. Avoid using databases with too many edges as this can increase the visual chaos of the network.
- When annotating protein complexes and biological processes, bring nodes that participate in the same biological entity in close proximity and surround them with a colorful halo (see Figure 2B). We recommend moving the halo to the background so as not to obscure protein names or edge colors.

**RESOURCE AVAILABILITY**

**Lead contact**

Further information and requests for resources and reagents should be directed to and will be fulfilled by the lead contact, Mehdi Bouhaddou (bouhaddou@ucla.edu).

**Materials availability**

Plasmids used in this study are all available on AddGene.

**Data and code availability**

Additional data are available from the lead contact upon reasonable request. This study did not generate new code.

**ACKNOWLEDGEMENTS**

This work was funded by National Institutes of Health grant U19AI171110 (N.J.K. and M.E.), U19AI135990 (N.J.K.), U19AI135972 (N.J.K.), K99AI163868 (M.B.), an HIV Accessory and Regulatory Complexes (HARC) Collaborative Development Award (M.B.), and a Center for AIDS Research Pilot Grant (M.B.). Also by Defense Advanced Research Projects Agency (DARPA) Cooperative Agreement #HR0011-19-2-0020 (N.J.K.; the views, opinions, and/or findings contained in this material are those of the authors and should not be interpreted as representing the official views or policies of the Department of Defense or the U.S. Government). We acknowledge additional funding from the Laboratory for Genomics Research (LGR) Excellence in Research Award #133122P (N.J.K.), F. Hoffmann-La Roche (N.J.K.), Vir Biotechnology (N.J.K.), UCLA-Caltech Medical Scientist Training Program (NIGMS T32 GM008042) (S.F.B), David Geffen Medical Student Scholarship (S.F.B), and gifts from QCRG philanthropic donors (N.J.K.). We thank Lars Plate and colleagues Katherine M. Almasy and Jonathan P. Davies (Vanderbilt University, Nashville, Tennessee) for the gift of plasmids expressing SARS-CoV-2 Nsp3 fragments.

**AUTHOR CONTRIBUTIONS**

Conceptualization, P.K., N.J.K., and M.B.; research design and methodology, P.K., M.R.U., G.M.J., J.X, B.P, Y.Z., E.S, M.E, R.K, D.S, L.Z.-A., N.J.K., and M.B; data analysis and interpretation, P.K., and M.B.; writing – review and editing, P.K., M.R.U., G.M.J., Y.D., S.K.M., S.F.B., D.M.W., J.X, B.P, Y.Z., E.S., M.E., L.Z.-A., R.K., D.L.S., N.K., and M.B.

**DECLARATION OF INTERESTS**

The Krogan Laboratory has received research support from Vir Biotechnology, F. Hoffmann-La Roche, and Rezo Therapeutics. N.J.K. has a financially compensated consulting agreement with Maze Therapeutics. N.J.K. is the President and is on the Board of Directors of Rezo Therapeutics, and he is a shareholder in Tenaya Therapeutics, Maze Therapeutics, Rezo Therapeutics, and Interline Therapeutics. M.B. is a scientific advisor for Gen1e LifeSciences.

**FIGURE LEGENDS**

Figure 1. Western blot analysis of all 18 HIV proteins reproduced from Jäger et al.[8] Anti-FLAG Western blot analysis of cell lysates of all 18 HIV-SF proteins after being transiently transfected into HEK293 cells.

Figure 2: Visualization of differential protein-protein interactions. (A) Heatmap of virus-host protein-protein interactions[1]. Rows contain nucleocapsid (N) viral protein mutants from each SARS-CoV-2 variant of concern. Columns contain interacting host proteins. Colors map to the log2 fold change (log2FC) between mutant and wave 1 (W1) form. Red indicates increased binding to mutant, and blue indicates decreased binding to mutant. Black boxes represent significant (p < 0.05) changes. White-dashed boxes indicate a prey detected in only the mutant (red) or W1 (blue). (B) Network of differential virus-host PPIs. Only significantly different interactions (|log2FC|>0.5 & p<0.05) are shown.

Table 1: Amino acid sequence of 2x-Strep tag for insertion at N- and C-terminus of protein.

Table 2: PolyJet / DMEM solution volume to create a master mix (Tube B).

Interaction Perturbations in Human Genetic Disorders. Cell *161*, 647–660. 10.1016/j.cell.2015.04.013.
16. Tsitsiridis, G., Steinkamp, R., Giurgiu, M., Brauner, B., Fobo, G., Frishman, G., Montrone, C., and Ruepp, A. (2023). CORUM: the comprehensive resource of mammalian protein complexes–2022. Nucleic Acids Res. *51*, D539–D545. 10.1093/nar/gkac1015.
17. Branon, T.C., Bosch, J.A., Sanchez, A.D., Udeshi, N.D., Svinkina, T., Carr, S.A., Feldman, J.L., Perrimon, N., and Ting, A.Y. (2018). Efficient proximity labeling in living cells and organisms with TurboID. Nat. Biotechnol. *36*, 880–887. 10.1038/nbt.4201.